\documentclass[12pt]{article}
\usepackage{cite}
\usepackage{amsmath,amsthm, amscd, amssymb, amsfonts}
\usepackage{appendix}
\usepackage{graphicx}
\usepackage{enumerate}
\usepackage{natbib}
\newtheorem{defi}{Definition}
\newtheorem{fallacy}{Fallacy}
\begin{document}
\title{Questioning the reasonableness of the quantum nonlocality debate\\
{\tiny
Universidad Nacional de Asunci\'{o}n,Ruta Mcal. J. F. Estigarribia, Facultad de Ciencias Exactas y Naturales, Km 11 Campus de la UNA, San Lorenzo-Paraguay}
}
\author{\small Justo Pastor Lambare\footnote{email: jupalam@gmail.com}}
\date{}
\maketitle
\begin{abstract}
We critically discuss the apparent lack of logical rigor pervading the debate on quantum nonlocality. Strong convictions often prevail over rational assessment, leading to the acceptance of loose ideas that become entrenched dogmas. The lack of sound rationales and adherence to the rules of logical inference lead to widely adopted antinomies that receive little conceptual scrutiny.
\end{abstract}
\tableofcontents  
\section{Introduction}\label{sec:Intro}
In this piece, we do not claim that every formulation of quantum locality or nonlocality is flawed.
Also, the intention is not to present a rigorous approach, nor to solve every interpretational issue.

The plan is to describe a curious situation in which puzzling, widespread contradictions and unsound interpretations are often presented as proofs that purportedly resolve a problem that is primarily interpretational and conceptual.
Back in 1982, baffled by this situation, the distinguished philosopher of science \cite{pFra82} observed: \textit{A reader as yet unfamiliar with the literature will be astounded to see the incredible metaphysical extravaganzas to which this subject has led}. 

In our opinion, calling ``metaphysical extravaganzas'' to the common inconsistencies and ad hoc fabricated interpretations is, in some cases, a euphemism.
In the first quarter of the 21st century, despite a Nobel Prize being awarded for the empirical falsification of the Bell inequality, the situation has not improved.

Nonlocality is identified with the idea of instantaneous action at a distance:
\begin{defi}\label{def:nl}
A theory is nonlocal if it predicts that what happens in a region of space can produce an immediate and instantaneous influence on another separate region without allowing a time delay for the propagation of the effect.
\end{defi}
We shall use the word ``instantaneous'' as a synonym for ``superluminal.'' 
Definition \ref{def:nl} is enough to convey Einstein's idea of ``spooky action at a distance'' in a pragmatic way.
The concept is theory-independent, i.e., applies equally to classical and quantum physics.
Thus, it makes perfect sense to ask whether or not quantum mechanics is a local theory.
Quantum mechanics is argued to be nonlocal because some of its predictions seem to satisfy Definition \ref{def:nl}.

Although Definition \ref{def:nl} represents the main idea behind the concept of nonlocality, we shall give another definition in Section \ref{sec:2dNONLOC} in terms of explanations, but the concept remains useful as long as it does not lose contact with the idea conveyed by Definition \ref{def:nl}.
Thus, although in other contexts it may be useful, in this discussion we avoid the use of terms such as ``Bell nonlocality'' and ``EPR nonlocality'', which suggest that certain kinds of nonlocality could exist (and they do) without implying the idea expressed in Definition \ref{def:nl}.

Another ubiquitous concept in the debate on nonlocality is realism.
The realism concept has been criticized for being ambiguous \citep{pNor07,pGis12,icLau12}; however, as usually understood by physicists in the nonlocality debate, it has a clear meaning   \citep{pCla78}:
\begin{defi}\label{def:real}
  Realism is a philosophical view, according to which external reality is assumed to exist and have definite properties, whether or not they are observed by someone.
\end{defi}
The realism tenet has its origin in the ``elements of physical reality'' concept introduced in 1935 by Einstein, Podolski, and Rosen (EPR) \citep{pEPR35}.
The EPR realism is just a contrived form of requiring  determinism since, by its very definition,
\begin{quotation}
  \textit{If, without in any way disturbing a system, we can predict with certainty (i.e., with probability equal to unity) the value of a physical quantity, then there exists an element of physical reality corresponding to this physical quantity.}
\end{quotation}
The relevant point is about predicting with probability equal to unity, which can be pragmatically called determinism.\footnote{In this discussion and following EPR's spirit, we do not distinguish between predictability with certainty and determinism as a more rigorous philosophical discussion perhaps requires \citep{pCav12}.}
The rest is metaphysical speculation about what purportedly exists without actually being observed.

It is worth noting that EPR's metaphysical speculation about physical reality is neither required by classical physics nor denied by the formalism of quantum mechanics.
Einstein's analytical mind immediately criticized the concept and never used it in his personal accounts \citep{pHow85}.
John Stewart Bell never invoked the concept either.

In Sections \ref{sec:NLandEPR}, \ref{sec:NandBT}, and \ref{sec:BIvsSI}, we review some widespread fallacies, sophisms, and unjustified interpretations surrounding the nonlocality problem.
Section \ref{sec:2dNONLOC} discusses two different concepts of nonlocality.
Finally, in Section \ref{sec:QLOC}, we go over some possible consistent arguments in favor of quantum locality.
%
\section{Nonlocality and EPR}\label{sec:NLandEPR}
The EPR paper is an argument for the incompleteness of quantum mechanics.
The core of the EPR argument for incompleteness proceeds as follows.

Let $LOC$ stand for locality and $R$ stand for realism, the EPR argument is a reasoning proving that,
\begin{equation}\label{leq:epr1}
  LOC \rightarrow R
\end{equation}
Conforming to (\ref{leq:epr1}), every local theory must contain elements of physical reality and comply with realism.
Quantum mechanics, being not deterministic according to the superposition principle and the Born rule, lacks elements of physical reality.
Since nature itself is assumed to be local and quantum mechanics does not comply with realism, it cannot be the whole story, so to avoid nonlocality, it is necessary to complete it with hidden variables.

\cite{pBoh35} responded to the EPR paper, rejecting the EPR's realism criterion, arguing that quantum mechanics is indeed complete.
According to the standard narrative, if $QM$ stands for quantum mechanics, Bohr's reply consists of the rejection of realism,
\begin{equation}\label{leq:Bor}
  QM\rightarrow \neg R
\end{equation}
The fallacy purportedly proving that quantum mechanics is a local theory because Bohr proved Einstein wrong is the claim that,
\begin{fallacy}\label{fal:fala1}
  Given the dichotomy between locality and realism, quantum mechanics is local because it is incompatible with the principle of realism.
\end{fallacy}
However, the purported dichotomy does not exist, or at least is not proved by those who claim it.
The concrete explanation of how the rejection of realism is supposed to restore locality to the predicted perfect correlation of distant events is absent.
The dismissive attitude seems to be ``Bohr somehow already explained all that.''

The irony of claiming locality by rejecting realism is that those concepts are logically unrelated unless connected by (\ref{leq:epr1}) through the EPR argument, in which case the correct logical inference confirms quantum nonlocality.

Indeed, since (\ref{leq:epr1}) is equivalent to $\neg R\rightarrow \neg LOC$, (\ref{leq:epr1}) and (\ref{leq:Bor}) prove exactly the opossite,

\begin{equation}
  QM\rightarrow \neg LOC
\end{equation}
Fallacy \ref{fal:fala1} is a contradiction since any proof of non-realism purportedly proving Einstein wrong, would be a proof of nonlocality.
Thus, by confirming that quantum mechanics is indeed ``not real'', Bohr would be indirectly confirming its nonlocality. 
Still, the apparent mainstream position seems to be that nonlocality is claimed only by EPR realists who, like Einstein, are unable to overcome their classical prejudices.

We remark, however, that reducing Bohr's response to (\ref{leq:Bor}) is an unjustified oversimplification that cannot be attributed to Bohr.
It would be puzzling that an intellectual giant like Niels Bohr committed such an obvious logical mistake, and his opponent, another giant of 20th-century physics, did not notice it.

Many consider Bohr's response to EPR as obscure, hard to understand, and even almost unintelligible.
For instance, \cite{pBel81} wrote: \textit{``While imagining that I understand the position of Einstein, as regards the EPR  correlations, I have very little understanding of the position of his principal opponent, Bohr.''}
Then Bell finished the section, wondering: \textit{``Is Bohr just rejecting the premise -- `no action at a distance' -- rather than refuting the argument?''}

Without us claiming a full comprehension of Bohr's response, it seems clear that it cannot merely be reduced to (\ref{leq:Bor}).
A more cogent interpretation would be that he rejected the very conditions under which the implication (\ref{leq:epr1}) was obtained.
At least that seems to have been Einstein's interpretation, who understood Bohr's response as denying the possibility of confronting what has been actually measured with what has not, calling Bohr's position solipsism.

Thus, we do not claim that Bohr's response reduces to Fallacy \ref{fal:fala1}.
The fallacy emerges in the perfunctory form in which Bohr's response is usually exploited in the nonlocality debate to the point that it has acquired the status of an effective ``orthodox dogma'' to banish any consistent or erratic nonlocality argument.

Fallacy \ref{fal:fala1} also arises in the context of the Bell theorem that we shall discuss separately.
%
\section{Nonlocality and the Bell theorem}\label{sec:NandBT}
%
The Bell theorem has proven to be a profuse source of misinterpretations and fallacies.
John Bell was an exceptionally clear-minded thinker, and he should not be charged for all the antinomies arising from misinterpreting his inequalities.
%
\subsection{Nonlocality and the ``deterministic'' Bell theorem}\label{ssec:NandDBT}
%
The first fallacy we analyze in this section is,
\begin{fallacy}\label{fal:BTfala1}
The 1964 Bell's theorem proves that quantum mechanics is nonlocal.
\end{fallacy}
An attentive reading of his paper \citep{pBel64} shows that Bell did not claim Fallacy \ref{fal:BTfala1}.
In \citep{pBel64}, Bell accepted the EPR reasoning of incompleteness from the beginning, as is clearly shown by part of the initial paragraph of his paper:
\begin{quotation}
  \textit{The paradox of Einstein, Podolski, and Rosen was advanced as an argument that quantum mechanics could not be a complete theory but should be supplemented by additional variables. These additional variables were to restore to the theory causality and locality.}
\end{quotation}
Thus, he was very explicit about his intention of inquiring into the admissibility of local hidden variables.
By accepting the EPR argument, he intended to ``restore locality'' to a theory that, being incomplete, was no longer local.
At least according to Bell, and contrary to how most non-localist advocates interpret it, his theorem is a no-go theorem for local-hidden-variables, as he again confirmed at the beginning of the conclusion section:
\begin{quotation}
    \textit{In a theory in which parameters are added to quantum mechanics to determine the results of individual measurements, without changing the statistical predictions, there must be a mechanism whereby the setting of one measuring device can influence the reading of another instrument, however remote.}
\end{quotation}
Clearly, he was declaring nonlocal any hidden variable theory\footnote{Rigorously, this is incorrect, but we are implicitly leaving out superdeterministic hidden variable models.} that preserves the predictions of quantum mechanics, not the theory of quantum mechanics itself. 
Of course, having accepted the EPR reasoning, for him, quantum nonlocality was not under discussion.
%
\subsection{Nonlocality and the ``stochastic'' Bell theorem}\label{ssec:NandSBT}
%
In 1975, Bell introduced the concept of local causality, which, unlike the EPR reasoning, is a locality condition for nondeterministic theories \citep{bBel04a}.

It is essential to understand the difference between EPR and local causality.
In the EPR case, locality requires determinism while local causality does not.
Local causality is a criterion allowing us to evaluate whether a nondeterministic theory is nonlocal without imposing on it the condition of becoming deterministic.

Only after arguing that quantum mechanics violates the local causality criterion did Bell turn to the question of whether it is possible to add hidden variables to restore locality, preserving the statistical predictions of quantum mechanics.
As in 1964, he found through his ``(stochastic) locality inequality'' that it was not possible to complete quantum mechanics without violating locality.

Despite the popular claim, by localists and non-localists, that Bell intended to prove quantum nonlocality with his stochastic inequality, he explicitly and unambiguously argued that quantum mechanics violates local causality before introducing his inequality and hidden variables on at least two occasions \citep{bBel04a,bBel90}.
So, there is sufficient textual evidence showing that Bell used his ``locality inequality'' only to investigate the admissibility of restoring locality by adding hidden variables and not to prove quantum nonlocality.
Thus, respecting Bell's formulation, we can state the next fallacy,
\begin{fallacy}\label{fal:BTfala2}
  The stochastic Bell inequality proves that quantum mechanics violates local causality.
\end{fallacy}

Since, as we explained above, Bell disassociated his ``locality inequality'' from his quantum nonlocality argument, we also have the next related fallacy,
\begin{fallacy}\label{fal:BTfala3}
  Quantum mechanics is local because the Bell theorem assumes realism.
\end{fallacy}
Fallacy \ref{fal:BTfala3} is doubly unjustified. 
First, as we explained above and respecting Bell's formulation, the Bell inequality can neither prove nor disprove quantum nonlocality.
Secondly, since the stochastic version of the inequality is not based on determinism, it is not at all clear how the elements of physical reality could be involved, unless we arbitrarily impose the ``realism dogma'' by misinterpreting the common causes $\lambda$ as pre-existing elements of physical reality.

Consistent counterarguments justifying quantum locality should dismiss the Bell inequality altogether and concentrate on rejecting the following concrete nonlocality arguments (cf. Sec. \ref{sec:QLOC}):
\begin{enumerate}[a)]
  \item The EPR argument, as explained in Sec. \ref{sec:NLandEPR}, and noticing that the mere naive rejection of realism as usually understood only confirms nonlocality.
  \item The violation of local causality by quantum mechanics, as argued by \cite{bBel04a}, or proved in \citep{ppLamb25}. In this case, the subterfuge of rejecting realism to justify locality is even more preposterous since local causality does not assume determinism.
      A consistent counterargument requires the discussion and eventual rejection of \textit{Reichenbach's common cause principle} \citep{bRei56}.

\end{enumerate}
%
%
\section{The Bell inequality vs. Stapp's inequality}\label{sec:BIvsSI}
%
Henceforth, we shall assume the Bell-CHSH inequality \citep{pCHSH69} as the Bell inequality.
\cite{pSta71}, realizing that the Bell inequality cannot be consistently considered a direct proof of quantum nonlocality, conceived a similar inequality to prove that quantum mechanics is indeed nonlocal.

Unlike the Bell inequality, Stapp's inequality was based on counterfactual reasoning and did not assume hidden variables.
While the Bell inequality is a no-local-hidden-variable result, the Stapp inequality is a genuine quantum nonlocality argument.

Table \ref{tabla:t1} compares the characteristics of both inequalities.
The Stapp inequality can be considered, along with the EPR reasoning and local causality, a third consistent argument for quantum nonlocality.\footnote{A detailed discussion of Stapp's inequality can be found in \citep{pLam22b}.}

Notwithstanding their different nature and purposes, the Stapp and Bell inequalities were soon incorrectly conflated and confused, spawning some of the wildest ideas existing in the nonlocality debate. 
\begin{table}[ht]
\caption{Bell vs. Stapp Inequality\label{tabla:t1}}
\begin{center}
\begin{tabular}{|c|c|c|c|c|}
\hline
Inequality & Hidden     & Quantum     & Quantum      & Counterfactual \\
  type     & variables  & nonlocality & completion   &    reasoning   \\
\hline
Bell       &   Yes      & No          &  Yes         &   No           \\
\hline
Stapp      &   No       & Yes         &  No          &   Yes          \\
\hline
\end{tabular}
\end{center}
\end{table}
\begin{fallacy}\label{fal:cr}
  The Bell inequality relies on counterfactual reasoning.
\end{fallacy}
Counterfactual reasoning requires the comparison of what has happened with what would have happened if some initial condition had been different.
That reasoning is present in the EPR argument and Stapp's inequality, incidentally, both arguments for quantum nonlocality.

The Bell inequality, being a completely different creature with a distinct purpose, is so stunningly straightforward that it looks suspicious, so some might want to upset it to make it look more convincingly sophisticated, and others perhaps to make it look untenable and easy to debunk.

The Bell inequality consists of the prediction of four different series of ``actually performed'' experiments $A_iB_k$ with four different setting combinations $i,k\in\{1,2\}$, which are randomly selected, nothing else! 
You don't need to worry about what-if scenarios; you only need to compare what has actually happened in different series of experiments.
The results obtained with the same settings are then averaged and finally summed,
\begin{equation}\label{eq:BIs}
  S=\langle A_1B_1\rangle +\langle A_1B_2\rangle +\langle A_2B_1\rangle - \langle A_2B_2\rangle
\end{equation}
The bound predicted for the series of experiments (\ref{eq:BIs}) depends on the theory used to make those predictions.
Nonconspiratorial hidden variable theories predict 
\begin{equation}\label{eq:BI}
|S|\leq 2 
\end{equation}
while quantum mechanics predicts $|S|\leq 2\sqrt{2}$, end of story.
The complicated aspects are left for experimentalists to empirically test the predicted results.
%
\subsection{Emergence of the counterfactual claims}\label{ssec:EofCFC}
%
When predicting four series of experiments, all of which are assumed to have been actually performed, there is no place for counterfactual reasoning or some concomitant idea such as counterfactual definiteness or incompatible experiments, despite abundant claims to the contrary \citep{pEbe77,pPer78,pHer78,pSky82,pDic93,pBla10,pWol15,pVirzi24,pHance24}.

The misunderstanding arises when the derivation of the Bell inequality begins incorrectly with (\ref{eq:ooCDF}), interpreting its terms according to Stapp's counterfactual rule, as defined in Definition \ref{def:SCR}.

Additionally, we will see that the misinterpretation also arises from confusing the algebraic steps of the correct mathematical derivation with the requirements of an actual experimental test.

For the sake of simplicity, we shall assume a deterministic hidden variable model.
According to this model, the predicted average for a series of experiments with setting $i,k\in\{1,2\}$ is
\begin{equation}\label{eq:ABavr}
  \langle A_iB_k\rangle =\int A_i(\lambda)B_i(\lambda)\,\rho(\lambda)\,d\lambda
\end{equation}
Leaving aside the difficulties and technical complexities of the real experiments, to test the inequality, all the experimentalist has to do is perform a series of experiments varying the two settings $i,k$ randomly and registering the results until they collect sufficient amount of data to evaluate the averages (\ref{eq:ABavr}) for each of the four possible setting combinations, and finally add the values to obtain (\ref{eq:BIs}).

Once it is clear what we have to do in the laboratory to obtain (\ref{eq:BIs}), we derive the hidden variable prediction by replacing (\ref{eq:ABavr}) in (\ref{eq:BIs}) and remembering that $A_i(\lambda)=\pm1$, $B_k(\lambda)=\pm1$ :
{\small
\begin{eqnarray}
  S &=& \int A_1(\lambda)B_1(\lambda)\,\rho(\lambda)\,d\lambda+\int A_1(\lambda)B_2(\lambda)\,\rho(\lambda)\,d\lambda\label{eq:icon}\\
    &=& +\int A_2(\lambda)B_1(\lambda)\,\rho(\lambda)\,d\lambda-\int A_2(\lambda)B_2(\lambda)\,\rho(\lambda)\,d\lambda\\
  S &=& \int\left[A_1(\lambda)B_1(\lambda)+A_1(\lambda)B_2(\lambda)+A_2(\lambda)B_1(\lambda)-A_2(\lambda)B_2(\lambda)\right]\rho(\lambda)d\lambda\label{eq:ooCDF_aux}\\
  S &=& \int\left[\,A_1(\lambda)[(B_1(\lambda)+B_2(\lambda)]+A_2(\lambda)[B_1(\lambda)-B_2(\lambda)]\,\right]\rho(\lambda)d\lambda\\
 |S| &\leq&  2\label{eq:fcon}
\end{eqnarray}
}
The claim that the Bell inequality requires counterfactual reasoning arises from confounding the algebraic steps (\ref{eq:icon})$\sim$(\ref{eq:fcon}) of the mathematical derivation with what has been actually done in the laboratory or what needs to be actually done to obtain (\ref{eq:BIs}).

All the experimentalist is supposed to do is to provide the numbers $A_iB_k$ obtained from measuring each particle of an entangled pair; nothing else or different should be expected or assumed about the experiment.

The individual integrals in (\ref{eq:icon}) represent the averages $1/N\sum_j (A_iB_k)_j$.
The $\lambda$ variable is a theoretical assumption that does not impose any restriction on how the experiment is supposed to be performed, except that the measured particles must come from the same entangled pair to generate each value $A_iB_j$.

The algebraic operations (\ref{eq:icon})$\sim$(\ref{eq:fcon}) in the derivation are allowed by the mathematical properties of numbers obtained as results of the experiment, i.e., they must be interpreted as steps done on data already received, which do not impose any restriction on how those numbers were obtained.
They should not be interpreted as experimental protocols necessary to obtain those results.

However, when in the algebraic step (\ref{eq:ooCDF_aux}), the expression in square brackets,
\begin{equation}\label{eq:ooCDF}
A_1(\lambda)B_1(\lambda)+A_1(\lambda)B_2(\lambda)+A_2(\lambda)B_1(\lambda)-A_2(\lambda)B_2(\lambda)
\end{equation}
is interpreted as a step the experimentalist must actually reproduce in the laboratory verbatim, instead of a mere arithmetic manipulation on data previously obtained, we immediately encounter an experimental conundrum.
As we shall see, this conundrum has led to declaring as real what does not exist (cf. Sec. \ref{ssec:CFD}) or to requiring several consecutive measurements on the same particle (cf. Sec \ref{ssec:IE}).

Although (\ref{eq:ooCDF}) has no meaning as an experimental protocol, that did not deter people from interpreting it as Stapp interpreted a similar expression in his inequality \citep{pSta71}:
\begin{defi}\label{def:SCR}
  Of these eight numbers, only two can be compared directly to experiment. The other six correspond to the three alternative experiments that could be performed but were not.       
\end{defi}
The former counterfactual interpretation makes perfect sense for Stapp's approach to his inequality; however, the counterfactual recipe for the falsifiable Bell hidden variables inequality is experimentally untenable and theoretically meaningless.

The concrete interpretation of expression (\ref{eq:ooCDF}) is that it represents data obtained in previous experiments that, by the commutative and associative properties of the arithmetic operations, were grouped as in (\ref{eq:ooCDF}) by the same value of $\lambda$.
This is possible because only a finite number of ``effective'' hidden variables exist for the Bell-CHSH experimental arrangement \citep{pLam21c}.

It is surprising how such naive confusion has become a widespread misunderstanding.
It is like claiming the results of a poll are invalid because it was presented in ascending age, while in the real poll, data were collected randomly.
Fortunately, statisticians and sociologists do not have to bear with the realism dogma.

The counterfactual misinterpretation of the Bell inequality has led to the following two inconsistent claims.
%
%
\subsubsection{Counterfactual definiteness}\label{ssec:CFD}
%
Although Counterfactual Definiteness (CFD) is supposed to be a rigorous technical term that expresses realism, its use is ambiguous, so it is necessary to clarify the meaning that we shall adopt in this discussion. 

Some use the term Counterfactual definiteness (CFD) as a synonym for counterfactual reasoning.
Others consider CFD a sort of EPR realism, but going even further regarding the properties of things that we do not observe but presumably exist\citep{wiki:cfd25}:
\begin{defi}\label{def:CFD}
    CFD is the ability to speak ``meaningfully'' of the definiteness of the results of measurements that have not been performed (i.e., the ability to assume the existence of objects, and properties of objects, even when they have not been measured).
    ......In such discussions ``meaningfully'' means the ability to treat these unmeasured results on an equal footing with measured results in statistical calculations. It is this (sometimes assumed but unstated) aspect of counterfactual definiteness that is of direct relevance to physics and mathematical models of physical systems and not philosophical concerns regarding the meaning of unmeasured results.
\end{defi}
There is at least one more connotation that seems to be somehow different, but all are related to some form of counterfactual reasoning.
Here we shall adopt Definition \ref{def:CFD}, which leads more conspicuously to the following fallacy,
\begin{fallacy}\label{fal:CFD}
    Violation of the Bell inequality proves counterfactual definiteness is false.
\end{fallacy}
The obvious reason the above statement is fallacious is that the Bell inequality has nothing to do with results that are not supposed to be ``actually'' measured. 
However, the assertion is so stunningly baffling that a few comments are in order.

We chose to use Definition \ref{def:CFD} because it more bluntly states the core of the problem: \textit{It is this (sometimes assumed but unstated) aspect of counterfactual definiteness that is of direct relevance to physics and mathematical models of physical systems and not philosophical concerns regarding the meaning of unmeasured results.}

So, by dismissing the conceptual analysis as ``philosophical concerns'' it assumes a \textit{``shut up and calculate''} attitude, declaring that once a mathematical expression is written, all conceptual problems become irrelevant.

This shut-up-and-calculate pragmatism proved to be very successful \citep{pKaiser14} and, in many cases, is well justified, but pursuing it without restrictions can lead to nonsense.
The situation reminds us of a famous quote attributed to Carl Sagan: \textit{``It pays to keep an open mind, but not so open your brains fall out.''}

For a moment, let us violate the shut-up-and-calculate dictum that Definition \ref{def:CFD} requires, and analyze expression (\ref{eq:ooCDF}) according to the CFD rule.
It claims that only one term corresponds to the pair that was actually measured. 
The remaining three correspond to the same pair of particles with the values that were not measured, but are real nonetheless,
\begin{equation}\label{eq:ooCDF1}
\underbrace{A_1(\lambda)B_1(\lambda)}_{actually\,\,measured}+\underbrace{A_1(\lambda)B_2(\lambda)+A_2(\lambda)B_1(\lambda)-A_2(\lambda)B_2(\lambda)}_{real\,\,but\,\,not\,\,actually\,\,measured}
\end{equation}
Note the difference with counterfactual reasoning, which is a mere prediction of what would have happened. 
However, CFD declares that what did not happen physically exists.
Let us ponder over what a Bell inequality violation would lead us to conclude under this hypothesis:
\begin{quotation}
    \textbf{Conclusion}: in view that the Bell inequality is violated in real experiments, according to (\ref{eq:ooCDF1}), we must conclude that what we did not actually measure does not exist.
\end{quotation}
Should we be surprised to find that what we did not measure does not exist?
Admittedly, perhaps to dissimulate this curious inference, Definition \ref{def:CFD} adds the caveat ``in statistical calculations.'' 
However, it does not significantly alter its meaning, except that the condition is relaxed, allowing the materialization to occur randomly after the first measurement, rather than immediately following it.

To further inquire into the reasonableness of subjecting the CFD hypothesis to an empirical test, assume now that we find the inequality is not violated, what would be our conclusion?
Would it mean that six-eighths of what we measured was not generated during the experiment, but somehow materialized from unrealized potentialities out of the remaining two-eighths that were actually measured?

In all Bell-type experiments, including the celebrated 2015 loophole-free ones \citep{pAsp15}, all the measurements performed in the experiments correspond to entangled pairs that were painstakingly generated by the experimentalists during the experiment, and there was no possibility that they could be ghost particles that pop out of thin air as materialization of ethereal possibilities that did not take place.

What is even more perplexing is that CFD claims that several different physical magnitudes (spin) physically exist in several (infinite) directions simultaneously, for instance \cite{pBou17} explains:
\begin{quote}
\textit{Bell's theorem clearly makes use of counterfactual definiteness; his inequality involves the correlations of the spins of the two particles in each of two different directions that correspond to non-commuting spin components. This use of counterfactuals is entirely appropriate because it is used to investigate a test for classical hidden variable theories.}
\end{quote}
It is rather ironic that in classical physics, we never observe an object having two different angular momenta simultaneously; in any case, such a claim would be more akin to quantum superposition.

Although CFD yields the correct mathematical result (\ref{eq:BI}), it is logically and conceptually disconnected from both the Bell and Stapp inequalities and even more so from any empirical test.
It rests on an untenable hypothesis that leads to a mystifying conclusion, which is nonetheless widely accepted only because it is believed to disprove EPR realism, thereby purportedly solving the quantum nonlocality problem. 
The entire CFD argument can be considered a higher-order fallacy because, being itself fallacious, it is conceived to support another fallacy, namely, that nonrealism disproves nonlocality.
%
\subsubsection{Incompatible experiments}\label{ssec:IE}
%
A related deviant interpretation associated with expression (\ref{eq:ooCDF}) is that it purportedly requires the measurement of incompatible observables \citep{pMuy94,pNie19,pVirzi24}.

Any argument requiring the simultaneous measurements of incompatible quantities contradicts orthodox quantum mechanics and should be immediately dismissed without further ado, be it the Bell inequality or the Stapp inequality.

We could blame Bell and Stapp for the heresy of daring to claim quantum nonlocality. 
Yet it would be unjustified to accuse them of having assumed the simultaneous measurements of incompatible observables or, relatedly, requiring a sequence of measurements on the same particle.

We surmise that the only way to come up with such a claim is the misinterpretation of (\ref{eq:ooCDF}) that, by the way, appears in both inequalities, but in different contexts.

The first such claim that we know of was made in 1972 \citep{pDCB72}.
At the request of an editor, Bell succinctly commented on the criticism  \citep{pBel75a}:
\begin{quotation}
\textit{The objection of de la Pe\~{n}a, Cetto, and Brody is based on a misinterpretation of the demonstration of the theorem......But by no means. We are not at all concerned with sequence of measurements on a given particle, or of pairs of measurements on a given pair of particles. We are concerned with experiments with which for each pair the ``spin'' of each particle is measured once only.}
\end{quotation}
We are afraid that Bell's brief explanation to De La Pe\~{n}a et al. was not explicit enough to prevent it from becoming a persistent misinterpretation and endemic problem for correctly assessing the significance of the Bell inequality.

Recently, some researchers had the original idea of using weak measurements to avoid the collapse of the wave function and perform more than one measurement on the same particle \citep{pVirzi24}, as suggested by de la Pe\~{n}a, Cetto, and Brody.
Although the experimental setup is remarkable, and we ignore the interpretation and possible applications of such experiments, we are certain that they do not test the Bell inequality.

However, the explanation given by \cite{pG&P25a} to justify the two sequences of measurements on the same particle was the use of the quantum identity,
{\small
\begin{equation}\label{eq:cqi}
  \langle A_1B_1\rangle +\langle A_1B_2\rangle +\langle A_2B_1\rangle - \langle A_2B_2\rangle= \langle A_1B_1 + A_1B_2 + A_2B_1 - A_2B_2\rangle
\end{equation}
}
The l.h.s of (\ref{eq:cqi}) is what Bell-type experiments actually measure, while the expression on the r.h.s,
\begin{equation}\label{eq:rqi}
  A_1B_1 + A_1B_2 + A_2B_1 - A_2B_2
\end{equation}
was measured in the experiment reported in \citep{pVirzi24} through successive weak measurements on the same pair of entangled particles.
So, they claim, they indeed were testing the l.h.s of (\ref{eq:cqi}) that represents the usual Bell-CHSH experiment.

\cite{pBel66} noted that an identity such as (\ref{eq:cqi}) is a quite peculiar property of quantum mechanics.
The sum of incompatible observables (\ref{eq:rqi}) represents an observable that is none of the individual terms in (\ref{eq:rqi}) and requires a completely different experiment that can not be performed by measuring its terms individually.
Bell presented the following example:
\begin{quotation}
  \textit{A measurement of a sum of [two] noncommuting observables cannot be made by combining trivially the results of separate observations on the two terms - it requires quite a distinct experiment. For example, the measurement of $\sigma_x$ for a magnetic particle might be made with a suitably oriented Stern-Gerlach magnet. The measurement of $\sigma_y$ would require a different orientation, and of $(\sigma_x + \sigma_y)$ a third and different orientation.}
\end{quotation}
Thus, the sum of the four incompatible observables in the r.h.s of (\ref{eq:cqi}) represents a single observable that is different from each of its terms.
Therefore, despite the use of weak measurements, the following fallacy persists,
\begin{fallacy}
    The Bell inequality assumes the simultaneous measurement of incompatible observables or successive incompatible measurements on the same particle.
\end{fallacy}

%
\section{Two different concepts of nonlocality}\label{sec:2dNONLOC}
%
Here we rehearse an explanation of why most non-localists claim that Bell's ``locality inequality'' is proof of quantum nonlocality, contradicting Bell's own understanding of his inequality as proof of a no-go theorem for local hidden variables.

A careful reading of Bell's manuscripts reveals that he conceived a different concept of nonlocality than most non-localists employ.
Bell's nonlocality concept is the following,
\begin{defi}\label{def:Bnl}
  \textbf{Explanatory nonlocality}: A physical theory is non-local when it lacks a local explanation for its predicted perfect correlations.
\end{defi}
Probably, Bell did not state the nonlocality directly in these terms; however, the expression consistently reflects Bell's local causality concept. 
In particular, he explicitly said that the correlations cry out for explanation \citep{bBel04a}.

It is essential to understand that the required explanation has to satisfy some criterion of locality; for instance, we can trace back the cause of the perfect singlet correlation to the common quantum state preparation at the source.
However, this explanation fails the EPR criterion and also the common cause principle. 

In the case of the EPR reasoning, as explained in \citep{pBel64}, when an arbitrary setting choice in laboratory $A$ instantaneously produces a different result in a distant laboratory $B$, to avoid action at a distance, the outcomes must have been already predetermined.
Note Bell's careful use of the word ``predetermined'' (determinism) instead of ``preexisting,'' thus avoiding the infamous ``elements of physical reality'' metaphysical speculation.

In the case of local causality, the requirement of determinism is excused by requiring a common cause explanation \citep{bRei56}.
Such an explanation is also absent in quantum mechanics \citep{bBel04a,ppLamb25}.

Einstein and Bell's position was that since no local explanation exists within the theory, it becomes nonlocal, and the explanation must be looked up outside the theory to avoid the existence of ``spooky'' influences.

Except for a few exceptions, we are not aware of a consistent counterargument given by advocates of locality. Fallacy \ref{fal:fala1} seems to be the universal locality argument wielded by localists.

According to Bell's (and Einstein) concept of nonlocality, the fact that a physical theory is nonlocal does not imply that nature itself is, unless it is impossible to add hidden variables either preserving its indeterminism (local causality) or turning it deterministic (EPR).

On the other hand, non-localists who assert that quantum mechanics is nonlocal because it violates the Bell inequality conceive of the concept of non-locality differently.
For them, a theory becomes nonlocal only after proving that it is not possible to add hidden variables to explain its correlations.

The above observations show that, by dismissing Bell's careful approach to the nonlocality issue, non-localist advocates exacerbate the polemic, inadvertently facilitating the realism dogma excuse.

Certainly, to a great extent, the radical disagreement and endless discussions between localists and non-localists are because non-localists dismiss Bell's Definition \ref{def:Bnl} of nonlocality, sidestepping the previous discussion of whether orthodox quantum mechanics is already non-local before adding hidden variables. 

Since, for localists, quantum mechanics itself is already local, nonlocality only emerges as the consequence of adding hidden variables that are foreign to quantum theory.
Ironically, the fact that hidden variables imply action at a distance was explicitly recognized by Bell in his 1964 paper's conclusion \citep{pBel64} (cf. the second Bell's quote in Sec. \ref{ssec:NandDBT}).

Finally, it is worth noticing that the correct nonlocality arguments (Einstein, Bell, Stapp) are disconnected from ontological assumptions about the nature of the quantum state or the wave function.
They involve only issues with the theory's objective predictions, such as the existence of perfect correlations.
The usual claim that the problem arises because of EPR realism is a sophism dogmatically fabricated only to reject the argument.

%
\section{Quantum locality}\label{sec:QLOC}
%
We have previously discussed three arguments for quantum nonlocality, namely, the EPR reasoning, local causality, and Stapp's inequality.
Here, we briefly mention what could be some consistent arguments for quantum locality. 

We begin by addressing counterarguments that reject the three bases of quantum nonlocality mentioned before.

The EPR reasoning and Stapp's inequality can be blocked by rejecting the possibility of counterfactual reasoning.
Bohr's reply to EPR goes in that direction by prohibiting the possibility of confronting the result of an experiment that we have performed with that which we have not.

Likewise, the violation of local causality by quantum mechanics can be dismissed by rejecting \textit{Reichenbach's common cause principle} \citep{bRei56}.
This is tantamount to rejecting that perfect correlations need an explanation and the rejection of causation.

Whether quantum correlations need explanations was considered by Bell, but he thought that: \textit{The scientific attitude is that correlations cry out for explanation} \citep{bBel04a}.
However, it is also possible to take the opposite position: \textit{So pervasive has been the success of causal models in the past, especially in a rather schematic way at a folk-scientific level, that a mythical picture of causal processes got a grip on our imagination} \citep{pFra82}.

A third argument for quantum locality could be to postulate that operational quantum mechanics is local. 
This position is empirically consistent thanks to the no-signaling property of quantum mechanics and, of course, accepting that correlations do not need a local explanation.

However, the problematic attitude of those subscribing to this operational position is that they usually ``want to have their cake and eat it too,'' meaning they do not want to concede that a local explanation indeed does not exist, pretending that the realism dogma explains everything.
A clear conceptual position must distinguish between accepting that an explanation is not needed and pretending that a trivial one exists (realism dogma).
%
\section{Conclusions}\label{sec:Conclu}
%
We have exposed several far-fetched fallacies and inconsistencies plaguing the quantum nonlocality debate, proposing instead, as mere down-to-earth examples, some rational interpretational alternatives.

The situation may be partly a consequence of a stagnation of critical thinking due to the unrestricted application of the ``shut up and calculate'' tradition, which, despite its heuristic value, should not eradicate rational thought.

Part of the peculiar situation is also due to the lack of clear definitions of fundamental concepts, such as quantum nonlocality, and the central role given to rather irrelevant metaphysical tenets, including EPR realism and counterfactual definiteness, which are devoid of any empirical substance and are eventually adopted as dogmas.
%
%
\bibliographystyle{chicago}
\bibliography{zBellbibfile}
\end{document}